\documentstyle[prl,aps,epsf]{revtex}
\makeatother
\def\abstract{\par
\ifpreprintsty %
\vskip2.5pc
\begin{center}%
{\large \abstractname\par}%
\end{center}%
\vskip.5pc
\fi
\bgroup
\ifdim\prevdepth=-1000pt \prevdepth0pt\fi
\hsize\columnwidth
\hsize2\columnwidth
\if@twocolumn\else\leftskip=0.10753\textwidth \rightskip\leftskip\fi
\dimen0=-\prevdepth \advance\dimen0 by17.5pt \nointerlineskip
\small\vrule width 0pt height\dimen0 \relax
}
\def\endabstract{\par\egroup}
\makeatletter

\begin{document}


\title{Creep motion in a granular pile exhibiting steady surface flow}
\author{Teruhisa S. Komatsu}
\address{Department of Environmental Science,
Japan Atomic Energy Research Institute, Tokai, Naka, Ibaraki 319-1195, JAPAN}
\author{Shio Inagaki, Naoko Nakagawa}
\address{
Department of Mathematical Sciences,
Ibaraki University,
Mito, Ibaraki 310-8512, JAPAN}
\author{Satoru Nasuno}
\address{
Department of Electrical Engineering, Kyushu Institute of Technology,
Tobata, Kitakyushu 804-8550, JAPAN
}
\twocolumn[
\maketitle
\begin{abstract}
We investigate experimentally granular piles exhibiting steady surface flow.
Below the surface flow, it has been believed exisitence of a `frozen' bulk region,
 but our results show absence of such a frozen bulk.
We report here that even the particles in deep layers in the bulk
 exhibit very slow flow
 and that such motion can be detected at an arbitrary depth.
The mean velocity of the creep motion
 decays exponentially with depth, and the characteristic
 decay length is approximately equal to
 the particle-size and independent of the flow rate.
It is expected that the creep motion we have seeen is observable
 in all sheared granular systems.
\end{abstract}
\ \\]

Granular materials exhibit behavior not seen
 in ordinary fluids or solids.
Considerable efforts have been made for understanding them,
 but further studies of various phenomena in granular materials
 are still required \cite{jn92,corsica,powder97,sm00}.
One of the typical behaviors of granular materials is 
 avalanche behavior of a sand pile \cite{jn92,bcpe94,degennes}.
In contrast to ordinary fluids,
 granular materials can form piles with a sloped surface.
When the angle of the surface exceeds some critical value,
 the pile cannot sustain the steep surface and an avalanche occurs.
Without careful observation, flow of particles in an avalanche appears
 to be limited to a surface layer, with a `frozen' bulk region below.
Many studies have been made under such an assumption
 without convincing experimental evidence \cite{jn92,bcpe94,degennes}.
Indeed, it is difficult to detect the frozen bulk clearly 
 because an avalanche is a transient behavior.
If such a frozen region actually exists and is well separated from
 the flow layer, it should be possible to observe and definitively
 identify it even in the case of steady surface flow.
With the purpose of making such an observation, we experimentally 
 investigated particle movements in piles exhibiting steady surface flow.

Here we report unexpected experimental findings in such piles:
Even the particles in deep layers are not frozen
 but exhibit very slow flow (creep),
 and such motion can be detected at arbitrary depth (Fig.1).
The mean velocity of the creep motion
 decays exponentially with depth, and that the characteristic
 decay length is approximately equal to
 the particle-size and independent of the flow rate.
This velocity profile differs from that for the surface flow,
 and it is in this sense that the flow
 of particles is separated into two regions.
We believe that creep motion of the type we have seen
 should be observable in all sheared granular systems.

Our experiments were performed in a quasi-two
 dimensional system, as described by Fig.2.
Alumina beads of diameter $a$ were continuously fed into the gap (A)
 existing between two vertical, parallel plates separated by width $W$.
In this gap, the particles formed a triangularly shaped pile with rapid flow
 on the surface slope, pouring out of the system from the right side (B).
The existence of the short wall on the right side eliminates
 the slipping of particles on the bottom of the cell.
Its presence thus assures that the motion of particles
 we observe is strictly due to the flow on the slope of the pile.
In order to maintain a steady surface flow on the pile,
 particles were continuously fed onto the pile from the left
 side using a ``double hopper'', in which the lower hopper was 
 continuously filled with particles by the upper one.
The mouth of the lower hopper was placed
 in such a manner that it always contacted the top of the pile.
This configuration allowed us not only to minimize
 the effect of the impact on the pile exerted by newly added
 particles but also to control the feeding rate by changing
 the height of the mouth of the lower hopper.
With this configuration, the feeding rate was essentially controlled
 by the flow on the pile, and it could thus be closely matched to
 the outflux from the system.
In this way a steady-state system could be established.
The flow rate $Q$ for each experimental run was determined by measuring
 the weight of the particles that poured from the right side of the gap
 per unit time.
With this system, the important control parameters are the
 particle diameter $a$, the flow rate $Q$, and the width of the gap $W$.
In the present study, we report the results obtained for three different
 diameters, $a=1.1\pm 0.1, 2.1\pm 0.1,$ and $3.1\pm 0.1$mm,
 and the gap width, $W=10a$.
We confirmed that qualitatively the same results are obtained
 for $W/a=5-40$.

As is seen in Figs.1,
 the surface of the pile is flat in the steady flow state, and
 its angle $\phi$ with respect to the horizontal direction
 is close to that after the flow stops when the supply of particles
 is cut off, i.e., the angle of repose.
In this state, the mean velocity is approximately parallel
 to the surface, and its functional dependence on the depth $h$
 (measured perpendicularly to the surface
 see Fig.2) is the same everywhere,
 except in the vicinity of left and right boundaries.
Figure 3 displays the velocity profiles as functions of
 the depth $h$ on a semi-log scale, while the inset displays
 the same data on a normal scale.
{}From the graph in the inset, the mean velocity $\langle v(h)\rangle$
 appears to decay as the distance from the surface increases, 
 and it appears to vanish at some finite depth.
Although in previous studies it has been assumed that
 these deep layers are `frozen',
 we found that the mean velocity is finite at all depths.
We call the slow motion of particles in such deep region as the {\it creep motion}.

As is clearly shown in Fig.3,
 for deep layers (large $h$) exhibiting creep motion,
 the velocity profile assumes the form of simple exponential decay:
\begin{equation}
\langle v(h)\rangle=v_0 \exp\left(-h/h_c\right),
\label{eq:exp}
\end{equation}
 where $h_c$ is the characteristic length.
While $v_0$ increases with the flow rate $Q$,
 $h_c$ is approximately equal to the particle diameter $a$
 for all values of $Q$.
The value of $h_c$ suggests that the creep motion is
 driven by events occuring on a particle-size scale.

Each velocity profile in Fig.3 is seen to bend at some depth.
At such a depth, there approximately exists a boundary at which
 the particle motion changes from rapid surface flow to slow creep motion.
As shown schematically in Fig.4,
 this boundary which is parallel to the surface,
 lies just above the upper edge of the right side wall.
The value $h_0$ corresponds to the thickness of the surface flow,
 which increases with the flow rate $Q$.
The particles below the boundary are jammed tightly
 due to the fixed wall existing below the surface downstream,
 while the upper surface flow is free from such obstruction.
The mean velocity $\langle v\rangle$ was found to have a longer
 characteristic length for the rapid surface flow above the boundary
 than for the creep motion;
 i.e, the creep motion decays more rapidly as a function of $h$
 than the surface flow.

In the region of the creep motion, particles are jammed tightly.
For such particles to move, the existence of voids is essential.
In fact, local rearrangements of particles are often observed
 around such voids.
In the creep region, we also observed plastic global deformation
 of the network formed by inter-particle connections.
Voids are created occasionally by such deformation,
 while this deformation is maintained by shear stress
 resulting from the surface flow.

These observations, together with the realization that the
 characteristic length of the decay is on the order of the particle size,
 lead to the simple idea that the exponential decay
 of the velocity profile can be understood
 in terms of an analysis based on the void creation process.
Let us now pursue this point.
For the region displaying creep motion we consider virtual layers
 parallel to the surface with particle size thickness.
Since the particles in these layers are severely obstructed,
 the mean velocity $v_n$ of the $n$-th layer
 (The index $n$ increases from the surface.)
 should be proportional to the production rate $P_n$ of voids:
 $v_n \propto P_n$.
A void in a given layer is generated by the escape of a particle
 or particles to neighboring layers,
 while this escape is driven by the relative motion
 between the neighboring layers.
Thus, we hypothesize that $P_n$ approximately satisfies
 $P_n \propto ( v_{n-1}-v_n )$.
This leads to
 $v_n = \alpha ( v_{n-1}-v_n )$,
 with some constant $\alpha$ ($>0$).
This ends us to expect that $v_n$ is an exponentially
 decaying function of depth.
We remark that these probabilistic considerations
 leave the unknown parameter $\alpha$.
Dynamical considerations, on the other hand, should allow us to
 determine a value for such a parameter,
 and this should provide an understanding of the physics of the granular
 materials behind this simple result.

We have reported only the results for alumina spheres.
Qualitatively, the same results were obtained for particles
 with different shapes and roughness:
 long seeds and irregularly shaped sand particles.
This suggests that our results do not depend sensitively on
 shapes nor roughness of particles.
The case of particles interacting through an attractive force and
 the case of heterogeneous particles are left as future problems.
We would like to briefly note the effect of creep motion on the surface flow.
Comparing the surface flow in presently investigated system with
 that on a truly frozen bulk
 (which can be modeled by a mono-layer of particles fixed
 on an inclined board),
 it is observed that flow in the former case is slower than that in the latter.
This implies that flow on a creeping bulk experiences greater resistance
 than that on a frozen bulk.
This is reasonable, because the surface flow itself is
 the source of the shear stress inducing the creep motion,
 and through this mechanism, energy is dissipated to the bulk below.
We believe that the slow creep behavior reported in this letter
 is common to all sheared granular systems, regardless of the source
 of the driving force, and that our results elucidate one of the fundamental
 properties of sheared granular systems.
While we studied steady states of granular piles,
 it is also interesting to consider the significance of
 our results with regard to unsteady phenomena,
 such as earthquakes \cite{scholz,marone,rubin}
 and the strengthening of granular piles \cite{nasuno,losert}
 exhibiting stick-slip phenomena under shear.
For example, the creep motion described by Eq. (1) might have some
 close relation with the logarithmic time dependences seen in these systems.

In summary,
 we have experimentally studied granular piles exhibiting steady surface flow.
We found that there exist rapid surface flow and
 slow creep motion below this flow.
The boundary between these two types of motion, across which the
 velocity profile changes continuously, is possible to be defined.
We do not find a `frozen' layer at any depth.
Rather, we found that the velocity of the creep motion decays exponentially
 according to Eq.~(\ref{eq:exp}) and that the characteristic length
 of this decay is on the order of the particle size
 and independent of the flow rate.
We believe that the behavior found here and the velocity profile
 we have observed are specific to granular materials.
We described a simple picture
 in terms of which to understand this velocity profile.
While most of our experiments were performed with spherical particles,
 we found that similar creep motion exists also for systems consisting
 of non-spherical particles.
This suggests that the behavior we have studied here is
 quite common (and perhaps universal) in granular systems.


The authors thank T. Sasaki of Ibaraki Univ. for his helpful advice
 and for allowing us to freely use his experimental equipment.
They also acknowledge S. Watanabe for his encouragement.
One of the authors (TSK) also acknowledges JSPS.
The analysis of captured images was performed using NIH Image.



\begin{figure}
\caption{
Snapshots of a granular pile in a steady flow state.
Particles were fed constantly from the surface of the left side,
 out of these figures.
The particles used were monodisperse alumina spheres
 with diameter $1.1\pm 0.1$mm (weight density $3.6$ grams/cm$^3$).
All photographs were taken under the same conditions,
 with only the shutter speeds differing;
(a)1sec, (b)1min, (c)1hour.
(a) A boundary between the surface flow and the `frozen' layers
 appears to exist at a depth of several particles below the surface.
(b) The apparent frozen layer of (a) is seen to flow.
Similarly, (c) reveals the flow of a thicker layer than that revealed by (b).
These observations reveal that the apparently `frozen' particles
 under the rapidly flowing surface are not stationary but slowly {\it creep}
 and that the layer in which this creeping motion can be detected 
 grows as observation time increases.
}
\end{figure}

\begin{figure}
\caption{
The geometry of the experiment.
We used a transparent material (acrylic plate) for the front plate
 (400mm wide) to allow observation of particle motion and aluminum for
 the back plate to reduce the static electricity effect.
The bottom and the left side of the gap were completely bounded
 by plates, whereas the right side was bounded only in its lower region
 (a region of height 20 mm).
}
\end{figure}

\begin{figure}
\caption{
Mean velocity profiles $\langle v\rangle$
 as functions of the depth $h$ from the pile's surface
 on a semi-log scale. The inset presents the same data on a normal scale.
 The horizontal axes repesent the depth from $h_0$
 normalized by the particle diameter $a$,
 where $h_0$ is the distance from the top of the fixed wall existing at 
 the right side of the system to the surface of the particles flowing over it.
 (See Fig.4.)
 The vertical axes represent velocity normalized by the value at $h_0$.
 The values of the particle diameter $a$(mm) and the flow rate $Q$(grams/sec)
 for the experiments are
 1 and 12 (squares),
 1 and 43 (solid squares),
 2 and 20 (circles),
 2 and 50 (solid circles),
 2 and 100 (triangles),
 3 and 34 (plus signs),
 3 and 58 (crosses).
 The broken line corresponds to $\exp(-0.72x)$.
}
\end{figure}

\begin{figure}
\caption{
A schematic picture showing the coexistence of the two different types
 of motion.
The boundary between them is parallel to the free surface,
 and its position seems to be determined by the height of the fixed wall
 existing downstream in our system.
}
\end{figure}

\end{document}